\documentclass{rmaa}
\usepackage{amsmath}

\title{Thermals in stratified regions of the ISM}

\author{A. Rodr\'\i guez-Gonz\'alez \& A. C. Raga
\affil{Instituto de Ciencias Nucleares, UNAM, M\'exico}}

\fulladdresses{
\item A. Rodr\'\i guez-Gonz\'alez, A. C. Raga:
Instituto de Ciencias Nucleares, Universidad
Nacional Aut\'onoma de M\'exico, Ap. 70-543, 04510 D. F., M\'exico
(ary,raga@nucleares.unam.mx)}

\shortauthor{Rodr\'\i guez-Gon\'alez \& Raga}
\shorttitle{Thermals in the ISM}

\SetVolume{40} \SetFirstPage{001} \SetYear{2013}
\ReceivedDate{\today} 
\AcceptedDate{Year Month Day} 

\resumen{Presentamos un modelo de una burbuja caliente que flota
dentro de una regi\'on del medio interestelar con estratificaci\'on
exponencial. Este modelo incluye t\'erminos representando el frenado
debido a la presi\'on hidrodin\'amica y debido a la incorporaci\'on
de masa ambiental a la burbuja. Calibramos los par\'ametros libres
asociados con estos dos t\'erminos mediante una comparaci\'on con
simulaciones tridimensionales de una burbuja que flota. Finalmente,
aplicamos nuestro modelo al caso de una burbuja caliente producida
por una supernova que explota dentro del medio interestelar
estratificado de la galaxia.}

\abstract{We present a model of a ``thermal'' (i.e., a hot bubble)
rising within an exponentially stratified region of the ISM. This model
includes terms representing the ram pressure braking and the entrainment
of environmental gas into the thermal. We then calibrate the free
parameters associated with these two terms through a comparison with
3D numerical simulations of a rising bubble. Finally, we apply our
``thermal'' model to the case of a hot bubble produced by a SN within
the stratified ISM of the Galactic disk.}


\begin{document}

\maketitle

\section{Introduction}

There are two simple models for buoyant flows in the Earth's
atmosphere:
\begin{itemize}
\item plumes: continuous flows produced by a heat source
in the base of the atmosphere,
\item thermals: rising bubbles resulting from an instantaneous
release of hot air.
\end{itemize}
These flows are described in detail, e.g., in the classic book
of Turner (1980).

Plumes have received considerable attention in the ISM literature,
having normally been called ``nozzles'' (Blandford \&
Rees 1974). These nozzle flows
resemble atmospheric plumes, but are fully compressible (plumes
in the Earth's atmosphere being in a highly subsonic, anelastic regime),
as discussed by Rodr\'\i guez-Gonz\'alez et al. (2009).

Less attention has been devoted to ``thermals''. Mathews et al. (2003)
and Nusser et al. (2006) develop ``hot bubble'' models for buoyant regions
within cooling flows in clusters of galaxies, including the effects of
the buouancy and the ``ram pressure braking'' of the bubble as it
moves within the surrounding environment.

Interestingly, both the astrophysical ``plume'' (nozzle) and ``thermal''
(``hot bubble'') models do not include terms representing the entrainment
of environmental material into the buoyant flow. Actually, Pope et al.
(2010) do discuss the possible importance of such a term, and evaluate
its effects on the dynamics of a rising hot bubble.
This situation is somewhat curious, since it is clear that the entrainment
term is fundamental for the development of buoyant flows in Earth's
atmosphere.

In the present paper, we develop a model for a fully
compressible (as opposed to anelastic) ``thermal'' (i.e.,
a positively buoyant ``hot bubble'') rising in an exponentially
stratified atmosphere, which includes terms representing the effects
of ``ram pressure braking'' and ``entrainment''. Both of these
terms lead to a slowing down of the motion of the bubble through
the environment (section 2). We then compare this model with 3D numerical
simulations of a buoyant bubble, and show that both the ram pressure
braking and the entrainment terms are necessary in order to
obtain a good agreement of the ``thermal'' model with
the numerical simulations (section 3).

Using the entrainment and ram pressure braking terms calibrated with
the numerical simulations, we then explore the full parameter space
of the ``thermal in an exponentially stratified atmosphere'' problem
(section 4). Also, we explore an application of this model to the case
of the hot bubble produced by a supernova exploding within the ISM
of the Galactic plane (section 5). Finally, the results are summarized
in section 6.

\begin{figure*}[!t]
\centering
\includegraphics[scale=0.5]{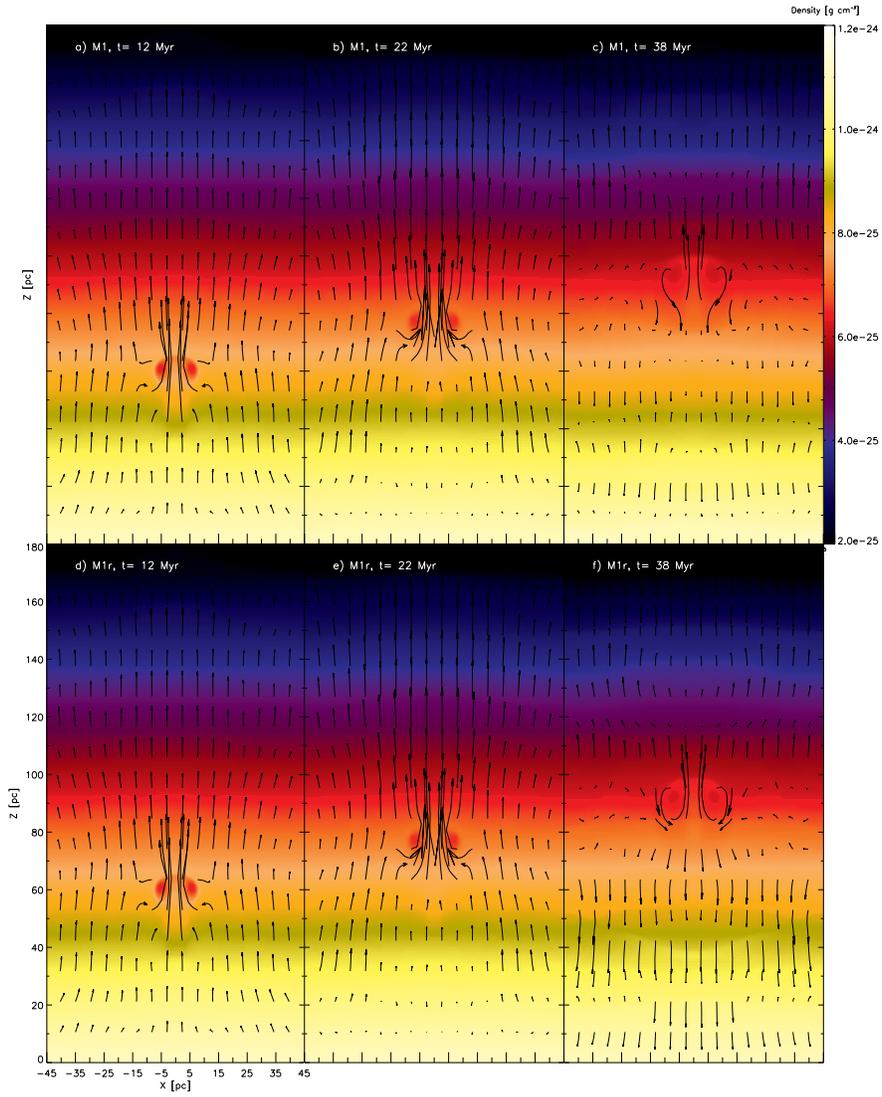}
\caption{Density stratifications on the mid $xz$-plane of the 3D
simulations of a rising bubble, obtained from models M1 (top) and M1r
(bottom frames). The initially spherical, hot bubble
is centered at $z_0=35$ pc, and the bubble then rises in the stratified
environment developing a ``ring vortex'' structure. This vortical structure
is traced by the flow velocity (shown with the arrows). Three frames are
shown, corresponding to $t=12$ Myr (left), 22 Myr (centre) and 38 Myr
(right) frames. The bar on the top right gives the density colour
scale in g cm$^{-3}$. From the displayed stratifications it is clear
that appreciable differences between the non-radiative (M1, top frames)
and radiative model (M1r, bottom frames) only appear at later
evolutionary times (i.e., for $t=38$ Myr).}
\label{fig1}
\end{figure*}

\begin{figure*}[!t]
\centering
\includegraphics[scale=0.7]{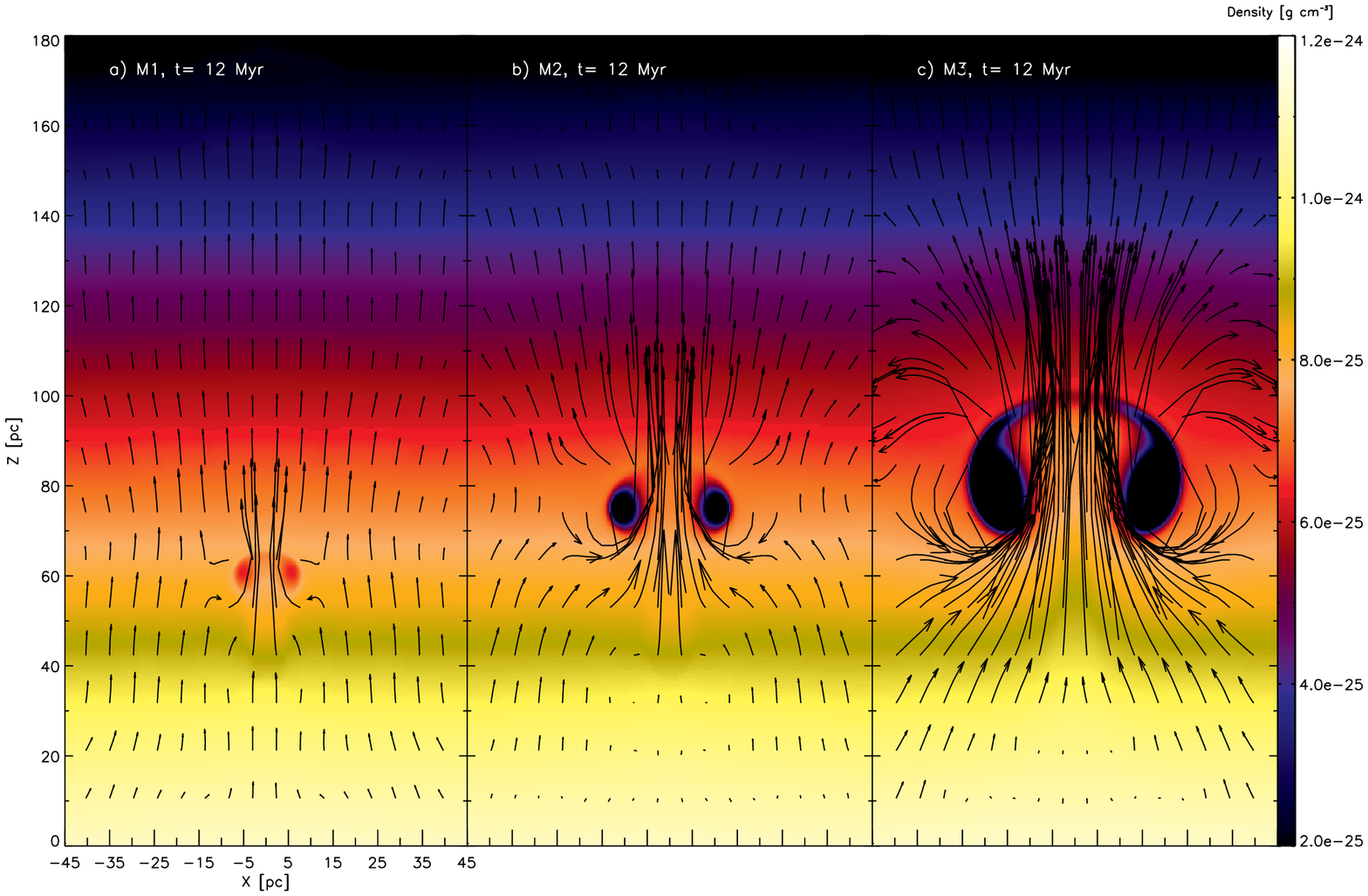}
\caption{Density stratifications and velocity fields
on the mid $xz$-plane of the 3D
simulations of a rising bubble, obtained from models M1, M2
and M3 for a $t=12$ Myr evolutionary time.
The bar on the right gives the density colour
scale in g cm$^{-3}$.}
\label{fig2}
\end{figure*}

\section{The compressible thermal model}

Let us consider a model of a hot bubble immersed in a cooler,
stratified environment with a pressure stratification $P_a(z)$
given by the hydrostatic condition:
\begin{equation}
\frac{dP_a}{dz}=-\rho_a g\,,
\label{hstat}
\end{equation}
where $\rho_a(z)$ is the ambient density stratification and $g$ is
the gravitational acceleration.

The pressure+gravitational force per unit volume on a parcel of
density $\rho$ in pressure
balance with the pressure $P_a$ of the surrounding environment then is:
\begin{equation}
f_z=-\frac{dP_a}{dz}-\rho g=\left(\rho_a-\rho\right)g\,,
\label{fz}
\end{equation}
where for the second equality we have used equation (\ref{hstat}).

We now write the equations for the time evolution of the ascending
parcel:
\begin{equation}
\frac{dM}{dt}=\rho_a v_e A\,,
\label{mass}
\end{equation}
where $M=\rho V$ (with $\rho$ the average density and $V$ the volume
of the parcel) is the mass of the parcel, $A$ is its surface area
and $v_e$ the so-called ``entrainment velocity'' associated with the
flux of ambient material incorporated into the volume of the ascending
parcel. The momentum equation is:
\begin{equation}
\frac{d\Pi}{dt}=\left(\rho_a-\rho\right)gV-\rho_a u|u|A_c\,,
\label{mom}
\end{equation}
where $\Pi=Mu=\rho Vu$ (with $u$ being the $z$-velocity) is the
momentum of the parcel along the $z$-axis. The second term on the
right hand side of this equation represents the drag due to
the ram pressure of the environment which is pushed aside by
the thermal, which presents an effective area $A_c$. This term is
negligible in the highly subsonic case, but becomes important for
mildly subsonic flows.

Finally, the energy equation can be written as:
\begin{equation}
\frac{dE}{dt}=-P_a\frac{dV}{dt}+\frac{P_a}{\gamma-1}v_e A\,,
\label{en}
\end{equation}
where $E=PV/(\gamma-1)$ is the thermal energy of the parcel
(assumed to be in local pressure equilibrium with the surrounding
environment) and $P_a/(\gamma-1)$ is the ambient thermal energy per
unit volume. As usual, $\gamma$ is the specific heat ratio (which
we assume has the same value for the parcel and for the ambient medium).
In this energy equation, we have assumed that the cooling due to
entrainment (second term in the right of equation \ref{en}) dominates
over the radiative cooling of the rising bubble.

We now close the system of equations (\ref{mass}-\ref{en}) assuming
that the ascending parcel has a homologous expansion (maintaining the
same shape) and that the entrainment velocity is proportional to the
velocity of the parcel, so that
\begin{equation}
A=V^{2/3}\,;\,\,\,\,\, v_e=\alpha |u|\,,
\label{ave}
\end{equation}
where the numerical factor in the $A/V$ relation is absorbed into
the $\alpha$ parameter (assumed to be constant) of the second relation
(as $A$ and $v_e$ always appear as the product $Av_e$, see equations
\ref{mass} and \ref{en}). We also write the effective area $A_c$
with which the thermal pushes away the environmental gas as:
\begin{equation}
A_c=\beta V^{2/3}\,,
\label{abeta}
\end{equation}
where $\beta$ is a constant of order unity.

With the relations of (\ref{ave}-\ref{abeta}), equations (\ref{mass}-\ref{en})
can then be written as:
\begin{equation}
{\rm mass:}\,\,\,\,\,\frac{dM}{dt}=\alpha \rho_a |u| V^{2/3}\,;\,\,\,\,\,
M=\rho V\,,
\label{mass2}
\end{equation}
$${\rm momentum:}\,\,\,\,\,\frac{d\Pi}{dt}=\left(\rho_a-\rho\right)gV
-\beta\rho_au|u|V^{2/3}\,;$$
\begin{equation}
\Pi=\rho Vu\,,
\label{mom2}
\end{equation}
\begin{equation}
{\rm entropy:}\,\,\,\,\,\frac{dS}{dt}=\alpha P_a |u| V^{\gamma-1/3}\,;
\,\,\,\,\, S=P_a V^\gamma\,,
\label{en2}
\end{equation}
where some simple manipulation has been made to convert (\ref{en})
into an entropy conservation form.

Finally, we specify a simple, plane, isothermal atmosphere
environmental stratification
\begin{equation}
\rho_a(z)=\rho_0\,e^{-z/H}\,;\,\,\,\,\,{\rm with}\,\,H=\frac{c_a^2}{g}\,,
\label{rhoa}
\end{equation}
which is the solution to equation (\ref{hstat}) for constant $g$. The
environmental pressure $P_a(z)=c_a^2\rho_a(z)$ (where $c_a$ is the
environmental isothermal sound speed) follows the same exponential law.

Equations (\ref{mass2}-\ref{rhoa}) can be integrated analytically
only in a partial way (see Appendix A).
We therefore integrate numerically equations
(\ref{mass2}-\ref{en2}) to obtain $V(t)$, $\rho(t)$, $u(t)$
and its time-integral
$z(t)$, for an exponential environmental density/pressure stratification
(with an isothermal sound speed $c_a$ and a gravitational
acceleration $g$, see equation \ref{rhoa}).
As initial conditions (at $t=0$) we set $z=0$, $u=0$
and choose an initial volume $V_0$ and density
ratio $\rho_c/\rho_0$ between the initial clump density
and the $z=0$ environmental density.

It is of course necessary to specify the two free parameters
$\alpha$ and $\beta$ (see equations \ref{ave}
and \ref{abeta}). As we have stated above, one expects to have
$\beta\sim 1$. Also, in order to reproduce laboratory experiments
of thermals it is well known that one needs to choose $\alpha\sim 0.1$.
In the following section we present 3D simulations of a thermal
in an exponential atmosphere. We then use a comparison of the
resulting $z(t)$ dependencies with an integration of equations
(\ref{mass2}-\ref{en2}) to determine the $\alpha$ and $\beta$ parameters.

\section{Numerical simulations}

In order to see whether or not the model presented in section 2 does
reproduce the features of a hot bubble rising in a stratified atmosphere,
we have computed 3D simulations of the flow. This has been done with
with a 3D version of the ``yguaz\'u-a'' code (Raga et al. 2000), using
a 4-level adaptive grid with a maximum resolution of $0.352$ pc
along the three axes. The domain has an extent of
$(90,90,180)$ pc along the $(x,y,z)$-axes, with reflection conditions
on the $\pm z$ boundaries and transmission conditions in all of the
other boundaries.

The equations for a $\gamma=5/3$ gas are integrated,
considering a gravitational force in the $-z$-direction (with constant $g$,
included in the momentum and energy equations). We have computed
non-radiative simulations, and simulations including the parametrized
radiative energy loss term of Raga \& Reipurth (2004), with
a low temperature cutoff at $10^4$~K.
The domain is initially filled with an isothermal, stratified density
structure $n_a=n_0\,\exp[-(z-z_0)/H]$ with $n_0=1$ cm$^{-3}$, $H=150$ pc
and $z_0=35$ pc.
This environment has a temperature of $10^4$ K (corresponding to
an isothermal sound speed $c_a=11.8$ km s$^{-1}$). The corresponding
value of $g$ is then computed as $g=c_a^2/H$.

The hot bubble is initially spherical,
located in the centre of the $xy$ range of the
computational domain, and at a height $z_0=35$~pc from the bottom
of the $z$-axis. The bubble has an initial density $n_c=0.01$ cm$^{-3}$
and temperature $T_c=10^6$~K, so that it is in pressure equilibrium
with the environment at a height $z_0$. We have computed non-radiative
and radiative models
using the three values of the initial radius $R_0$ of the bubble which
are given in Table 1.

Starting with this initial condition, the 3D Euler equations are
integrated forward in time, following the rise of the positively
buoyant bubble and its eventual mixing with the environment. Three
frames of the resulting time evolution of models M1 (non-radiative)
and M1r (radiative, see Table 1) are shown in Figure 1.

At a time $t=12$ Myr (left frames of Figure 1),
the bubble has risen to a height $z\approx 50$ pc,
and has developed a vortical structure, which is maintained throughout
the evolution of the flow. At $t=22$ Myr, the bubble has risen
to a height $z\approx 65$ pc, and the vortex ring has
expanded sideways quite considerably. At $t=38$ Myr (right frames
of Figure 1), the rise of the vortex has slowed down quite considerably,
reaching a maximum height of $\approx 95$ pc. At longer times the
vortex slowly drops to $z\sim 90$ pc, while mixing heavily with
the environment, and rapidly becoming unrecognizable
as a coherent structure.

From Figure 1, we see that the non-radiative (M1, top) and
radiative (M1r, bottom) simulations only show appreciable
differences at the later, $t=38$ Myr evolutionary time. Approximately
the same maximum height is attained by the thermal in the radiative
and in the non-radiative models.

Figure 2 shows the mid-plane density stratifications and flow
fields at $t=12$ Myr obtained from our 3 non-radiative models
(M1, M2 and M3),
illustrating the flow configurations obtained for the three
chosen initial radii of the thermal (see Table 1). At this
integration time, the flows obtained from the corresponding
radiative simulations (models M1r, M2r and M3r) are basically
identical to the corresponding non-radiative flows.

From the simulations, we can determine the height $z-z_0$ (where $z_0$ is
the height of the initial, hot bubble) as a function of time $t$. We
associate the (time-dependent) height of the ascending thermal with
the position at which the vertical velocity has its maximum value,
which approximately corresponds to the centre of the rising vortex ring.
The resulting height vs. time dependencies obtained
from our six simulations are shown in Figure 3.

\begin{figure}[!t]
\centering
\includegraphics[scale=0.7]{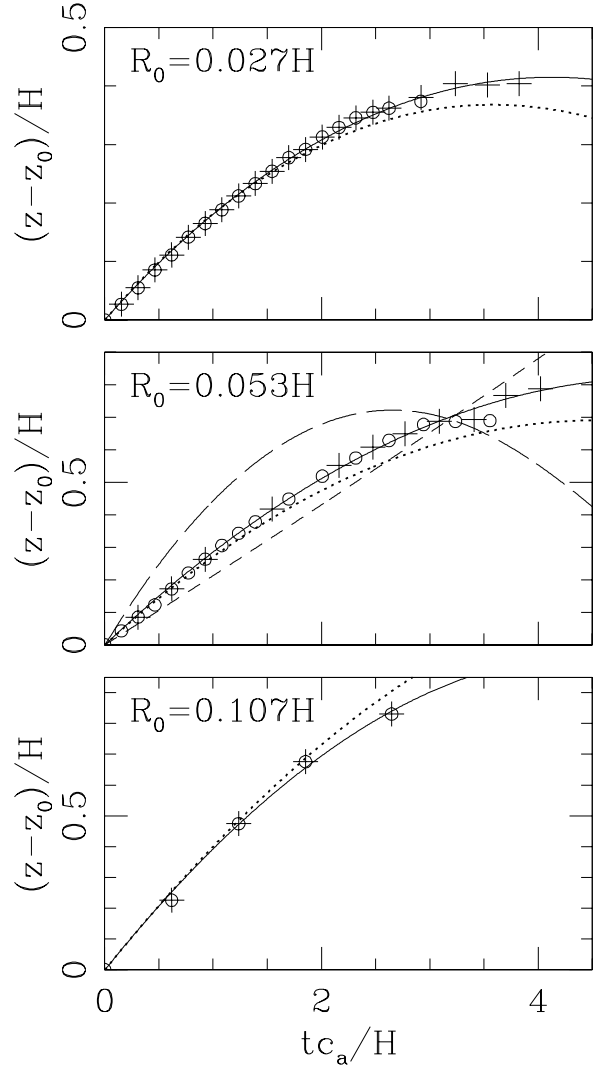}
\caption{Height of the hot bubble obtained from
the numerical simulations as a function of time (crosses: non-raditive
models, open circles: radiative models) and fits to these
time dependencies with ``thermal'' models (solid, dashed and dotted
lines). The top frame shows models M1 and M1r, the central frame
models M2 and M2r and the bottom frame models M3 and M3r.
The solid lines correspond to thermal models with the $\alpha$
and $\beta$ parameters corresponding to least squares fits
to the $z$ vs. $t$ dependencies obtained from the three
non-radiative simulations. The dotted lines correspond
to a thermal model with $\alpha=0.2$ and $\beta=0.7$ (the
average of the values obtained for the 3 non-radiative models,
see Table 1).
In the central frame, we have included the $z$ vs. $t$
dependencies obtained from best fit $\alpha=0$ (short dashes)
and $\beta=0$ (long dashes) thermal models, illustrating
the fact that in order to reproduce the simulations
it is necessary to include both the entrainment and
the ram pressure braking terms.}
\label{fig3}
\end{figure}

We then compute ``thermal'' models (see section 2) with an initial
bubble to environment density ratio $\rho_c/\rho_0=0.01$, and with
an initial volume $V_0=4\pi R_0^3$ (corresponding to the
parameters of the numerical simulations, see above
and Table 1), and with
arbitrary values of the dimensionless parameters $\alpha$ and
$\beta$ (see equations \ref{ave} and \ref{abeta}). Through
least squares fits to the (appropriately adimensionalized) position
of the rising vortex in the 3D simulations, we obtain the best
fit thermal models, shown with solid lines in the three frames
of Figure 3. The values of $\alpha$ and $\beta$ obtained
from fits to the three non-radiative models are given in
the two last columns of Table 1 (very similar numbers being
obtained from fits to the non-radiative models).

In order to show that it is actually necessary to have non-zero
$\alpha$ and $\beta$ values, we have also carried out least
squares fits to model M2 setting $\alpha=0$ (from which
we obtain $\beta=1.91$)
and $\beta=0$ (from which we obtain $\alpha=0.27$). These
best fits (shown with the dashed lines in the central
frame of Figure 3) fail to
reproduce the $z(t)$ dependence obtained from the M2 numerical
simulation in a satisfactory way.

From this comparison, we conclude that in order to reproduce the
rise of the hot bubble obtained from the numerical simulation,
the quasi-analytic ``thermal'' model needs to have both the ``entrainment''
and the ``ram pressure braking'' terms included in equations
(\ref{mass2}-\ref{en2}). The dimensionless parameters (associated with
these two terms) deduced from fits to the numerical simulations are
$\alpha\approx 0.2$ and $\beta\approx 0.7$, which correspond to
the average of the three values obtained by fitting models M1, M2
and M3 (see Table 1).

Needless to say, the numerical simulations which we are using have
a limited spatial resolution, and the values deduced for $\alpha$ and
$\beta$ will probably differ when changing the resolution of the simulation.
However, the fact that these two parameters are consistent with their
expected values indicates that the results that we obtained are approximately
correct.

\def\pz{\phantom{0}}
\begin{table}[!t]
  \newcommand{\DS}{\hspace{.7\tabcolsep}} 
  \setlength{\tabnotewidth}{\linewidth}
  \setlength{\tabcolsep}{1.0\tabcolsep}
  \tablecols{5}
  \caption{Radial collapse models}
  \begin{tabular}{ccccc}
    \toprule
Model\tabnotemark{(a)} & $R_0$\tabnotemark{(b)} &
$R_0/H$\tabnotemark{(c)} &
$\alpha$\tabnotemark{(d)} &
$\beta$\tabnotemark{(e)} \\
\midrule
M1, M1r & $\pz 4$~pc & 0.027 & 0.181 & 0.774 \\
M2, M2r & $\pz 8$~pc & 0.053 & 0.169 & 0.678 \\
M3, M3r & $16$~pc    & 0.107 & 0.278 & 0.606 \\
    \bottomrule
    \tabnotetext{a}{models M1-3 are non-radiative and models M1r-3r are
radiative}
    \tabnotetext{b}{initial radius $R_0$ of the bubble}
    \tabnotetext{c}{$R_0$ in units of the environmental scaleheight}
    \tabnotetext{d}{``entrainment'' parameter}
    \tabnotetext{e}{``ram pressure braking'' parameter}
  \end{tabular}
\end{table}

\section{Exploration of the parameter space}

We now consider the ``rising thermal'' model (see section 2) with
the $\alpha=0.2$ (entrainment) and $\beta=0.7$ (ram pressure drag)
parameters deduced from the comparison with 3D numerical simulations
(see section 3). If we write the height $z$ in units of the environmental
scale height $H$, the vertical velocity $u$ in terms of the environmental
isothermal sound speed $c_a$ and the time $t$ in units of $H/c_a$, the
problem has two free parameters: the initial volume $V_0$ (in units
of $H^3$) and the initial clump to environment density ratio
$\rho_c/\rho_0$.

Integrating numerically Equations (\ref{mass2}-\ref{en2}),
we have computed three ``thermal'' models with $\rho_c/\rho_0=0.1$
and $V_0=0.02$, 0.1 and $0.5\,H^3$. The resulting dimensionless
height $z/H$ and vertical velocity $u/c_a$ are plotted as
a function of time $t\,c_a/H$ in Figure 4. It is clear that
for increasing values of $V_0$ the rising thermal reaches
a larger maximum height, before falling again following a
damped, oscillatory behaviour.

\begin{figure}[!t]
\centering
\includegraphics[scale=0.7]{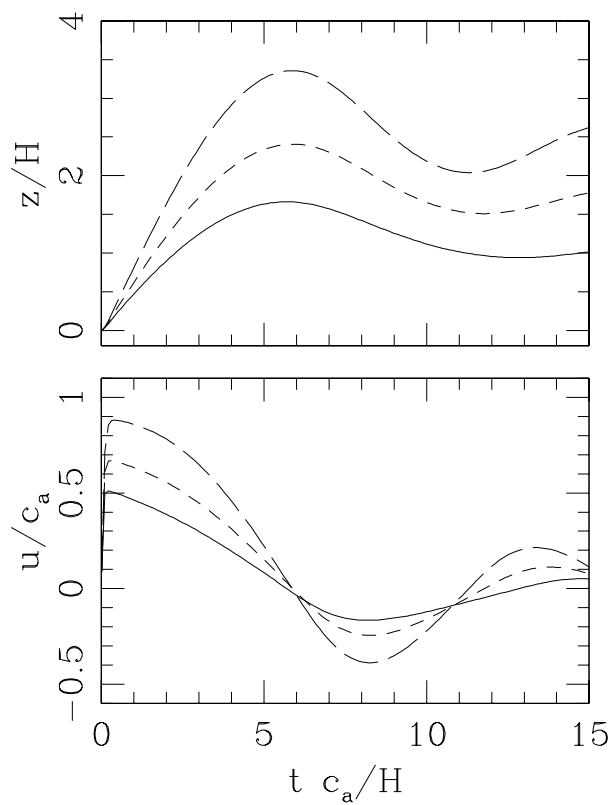}
\caption{Thermal model with $\alpha=0.2$, $\beta=0.7$,
$\rho_c/\rho_0=0.1$ and $V_0=0.02\,H^3$ (solid lines),
$0.1\,H^3$ (short dashes) and $0.5\,H^3$ (long dashes).
The height of the thermal as a function of time is shown on
the top frame, and its velocity in the bottom frame.}
\label{fig4}
\end{figure}

As we have shown in section 2, 3D numerical simulations produce
a rise which is similar to the one predicted by the quasi-analytic
``thermal'' model. However, once the maximum height is reached
the rising vortex obtained in the numerical simulations expands
laterally and mixes heavily with the environment, and only
shows a small drop in height from its maximum. Therefore,
the oscillatory behaviour obtained at larger times from the
single-parcel ``thermal'' model does not correspond to a real
physical phenomenon. Applications of the rising ``thermal'' model
therefore have to be limited to the initial rise of the hot bubble.

We have then computed a matrix of ``thermal''
models with $0< \rho_c/\rho_0\leq 0.5$
and $0< V_0 \leq 0.5\,H^3$. From these models we have computed
the maximum height $z_{max}$ reached by the rising thermal. The
resulting values of $z_{max}$ as a function of $\rho_c/\rho_0$ and
$V_0$ are shown in Figure 5.

\begin{figure}[!t]
\centering
\includegraphics[scale=0.45]{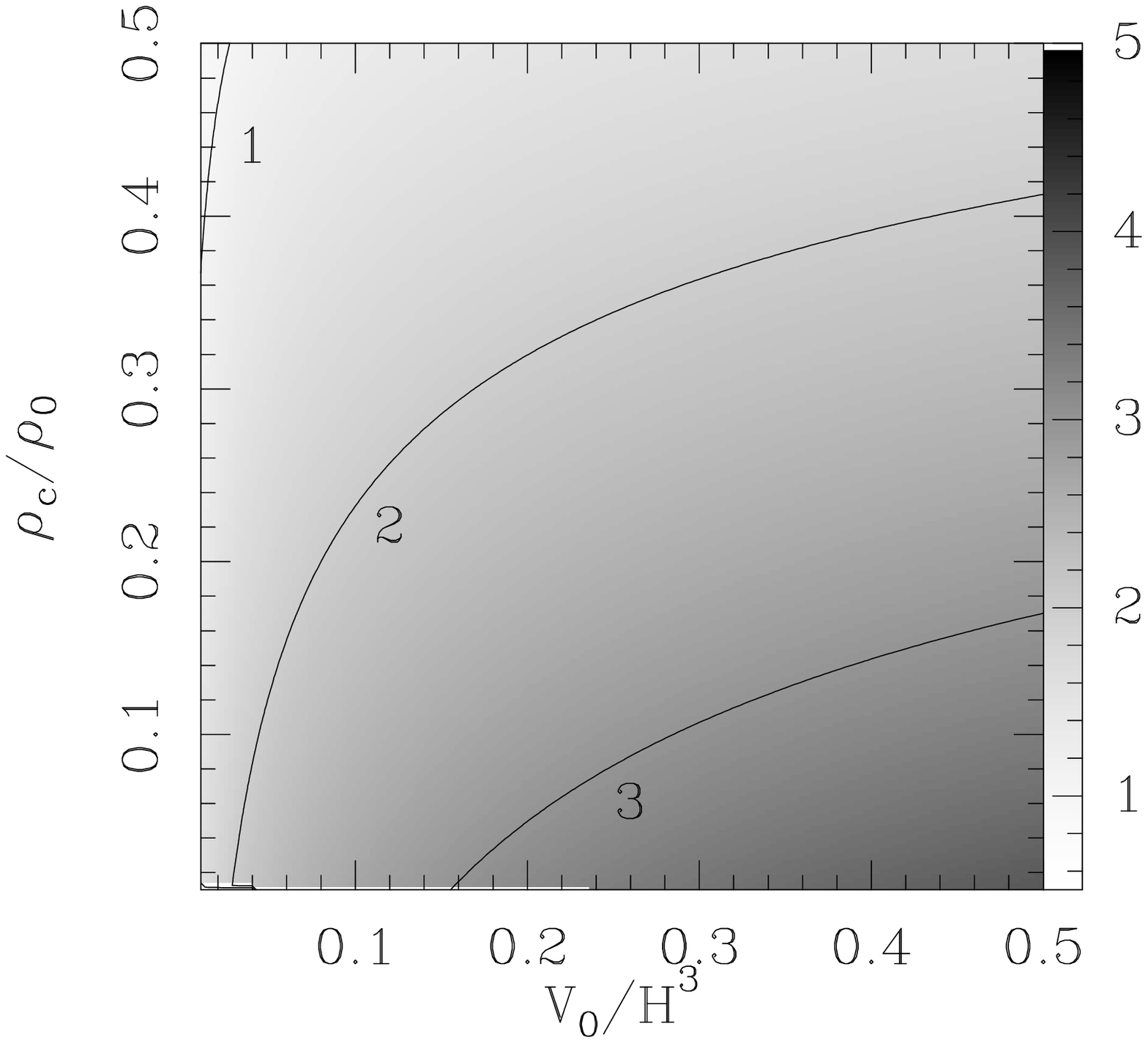}
\caption{Maximum height $z_{max}$ attained by the thermal as a function
of its initial volume $V_0/H^3$ (with $H$ being the environmental
scale height) and its initial clump to environmental density ratio
$\rho_c/\rho_0$. The labeled contours correspond to $z_{max}/H=1$, 2
and 3.}
\label{fig5}
\end{figure}

From this Figure, it is clear that for $V_0> 0.02\,H^3$
one has $z_{max}>H$ for all of the explored values
of $0< \rho_c/\rho_0\leq 0.5$. In order for the thermal
to reach $z_{max}\sim 3\,H$, one needs to have
low $\rho_c/\rho_0$ values ($\sim 0.1$) and high
values of $V_0$ ($\sim 0.3\,H^3$), corresponding
to the lower, right hand region of Figure 5.

\section{A supernova bubble within the warm ISM}

This model can be applied to the case
of a supernova (SN) explosion embedded in a dense region of
warm ($n\sim 10^3$ cm$^{-3}$, $T\sim 10^3$ K) gas in the plain
of the Galaxy. Raga et al. (2012) have shown
that the hot bubble produced by a SN explosion first
expands and then reaches a maximum radius
\begin{equation}
R_f=\left[\frac{3(\gamma^2-1)}{8\pi\gamma}\frac{E}{\gamma \rho_0c_a^2}\right]
^{1/3}\,,
\label{rf}
\end{equation}
in a time
\begin{equation}
t_f=\frac{(\gamma+1)R_f}{2\sqrt{\gamma}c_a}\,,
\label{tf}
\end{equation}
where $\rho_0$ is the density and $c_a$ the (isothermal)
sound speed of the uniform environment. At this evolutionary
stage, the hot bubble is in approximate pressure equilibrium
with the surrounding environment.

Let us now consider a SN explosion within a warm ISM region
of number density $n_0=1$ cm$^{-3}$
and sound speed $c_a=10$ km s$^{-1}$. For a SN energy
$E=10^{50}$ erg, from equations (\ref{rf}-\ref{tf}) we then
obtain $R_f\approx 81$ pc and $t_f\approx 8.2\times 10^6$ yr.
The temperature of this hot bubble is much larger than the
one of the cloud, so that the bubble to environment density
ratio will have a value $\rho_c/\rho_0\ll 1$.

As the warm ISM in the Galaxy has a scale height $H\approx 150$ pc,
the SN bubble has an initial radius $R_f\approx
0.54\,H$, and therefore an initial volume
$V_c=4\pi R_f^3/3\approx 0.66\,H^3$. We then compute
a ``thermal'' model with this value of $V_c$ and with
$\rho_c/\rho_0=10^{-2}$ (the actual value of this ratio
being unimportant provided that it is $\ll 1$). The
results from this model are shown in Figure 5.

\begin{figure}[!t]
\centering
\includegraphics[scale=0.7]{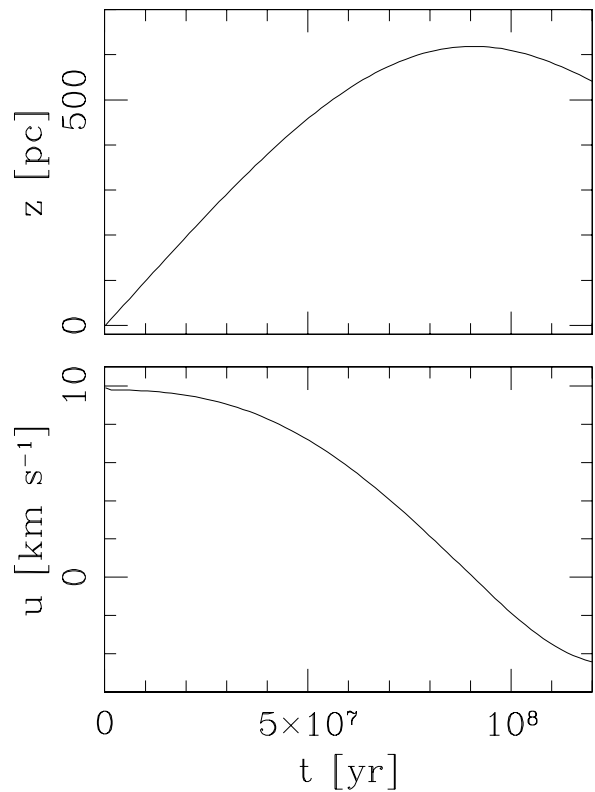}
\caption{Dimensional height (top) and velocity (bottom) for the
model of a SN bubble rising in the stratified, warm ISM of the
Galactic plane (see section 5).}
\label{fig6}
\end{figure}

From this Figure, we see that the initial bubble rises at
a transonic velocity of $\sim 10$ km s$^{-1}$
and then slows down, reaching a maximum
height $z_{max}\approx 600$ pc (of 4 environmental
scale heights) at a time $t\approx 90$ Myr. Therefore,
before reaching this point in reality the rising bubble
will emerge from the stratified ISM of the Galactic disk into
the hot ISM of the halo.

\section{Summary}

This paper describes a quasi-analytic ``rising thermal'' model: a
compressible, single parcel model including ``entrainment'' and
a ``ram pressure braking'' terms. This model differs from ``thermal''
models from the atmospheric sciences in that it is fully compressible
(atmospheric thermals being approximately anelastic) and in the presence
of the ram pressure braking term. This term is active in the transonic regime
relevant for astrophysical thermals, but is not important in the
highly subsonic, atmospheric case. We limit our study to the case of
a thermal in an exponentially stratified, isothermal atmosphere.

The model (see section 2)
has two free dimensionless parameters: $\alpha$, associated
with the entrainment term and $\beta$, associated with the ram pressure
braking term (see equations \ref{ave}-\ref{abeta}). From fits to
3D numerical simulations (see section 3), we find estimates
$\alpha=0.2$ and $\beta=0.7$ for these two parameters. We find
that the two terms (ram pressure braking and entrainment) are
necessary for reproducing the results obtained from the numerical
simulation.

With these $\alpha$ and $\beta$
parameters, we then compute a grid of models varying
the initial parameters $V_0/H^3$ (the ratio between the initial volume
of the hot bubble and the cube of the environmental scale height) and
$\rho_c/\rho_0$ (the initial ratio between the densities of the bubble
and of the surrounding environment), obtaining the maximum height
$z_{max}$ attained by the rising thermal (see section 4). We find
that for $V_0/H^3\sim 0.4$, one has thermals that rise several scale
environmental heights.

Finally, as a possible application of our model, we
describe the case of a SN explosion within
the warm ISM of the Galactic disk (section 5).
We consider the hot bubble produced by
an $E=10^{50}$ erg SN within an environment of density $n_0=1$ cm$^{-3}$
and sound speed $c_a=10$ km s$^{-1}$, with a scale height $H=150$ pc.
We show that the SN bubble takes $\sim 8$ Myr to reach a pressure equilibrium
radius $R_f\sim 80$ pc ($\approx 0.5\,H$). Over a longer timescale,
this hot bubble rises in the stratified ISM of the Galactic disk,
reaching a height of $\sim 600$ pc in $\sim 90$ Myr. The bubble will
therefore leave the disk ISM, and enter the hot ISM associated with the
Galactic halo.

Such rising thermals might be important for feeding the turbulence
of the Galactic ISM, and the models presented in this paper would
provide a clear guide for the future calculation of numerical
simulations of this process. Our present model could possibly also
be applied to other transient flows in the ISM in which buoyancy plays an
important effect. An example of such flows is of course the dynamics
of buoyant bubbles within cooling flows (see Pope et al. 2010 and
references therein).

\acknowledgments
We acknowledge support from the CONACyT grants
61547, 101356, 101975, 165584, 167611 and 167625, and the DGAPA-UNAM
grants IN105312 and IN106212.

\vskip1cm
\appendix{APPENDIX A: PARTIAL ANALYTIC SOLUTION OF THE THERMAL MODEL}
\vskip1cm

The system of equations (\ref{mass2}-\ref{en2}) for a rising thermal
can be integrated in a partial way. We start from the ``entropy conservation''
equation (\ref{en2}). Setting $P_a=c_a^2\rho_a$, with $\rho_a$ given
by equation (\ref{rhoa}) and using the relation $d/dt=u\,d/dz$,
equation (\ref{en2}) can be written in the form:
\begin{equation}
\frac{d}{dz}\left(V^\gamma e^{-z/H}\right)=\alpha V^{\gamma-1/3}
e^{-z/H}\,.
\label{en3}
\end{equation}
Now, setting $V^{\gamma-1/3}\approx V^\gamma/V_m^{1/3}$ (where $V_m$ is
the average volume of the rising parcel), equation (\ref{en3}) can
be integrated to obtain:
\begin{equation}
V(z)=V_0\, e^{z/H_1}\,;\,\,\,{\rm where}\,\,\,\frac{1}{H_1}=
\frac{1}{\gamma}\left(\frac{1}{H}+\frac{\alpha}{V_m^{1/3}}\right)\,.
\label{volz}
\end{equation}
One of course does not know the value of the average volume
$V_m$. We find that if we set $V_m=3V_0$ (where $V_0$ is the
volume of the thermal at $z=0$), equation (\ref{volz}) agrees
with the exact (i.e., numerical) integration of equation (\ref{en3})
to within 4\%\ in the $z=0\to 4H$ range.

We can then insert this $V(z)$ solution into equation (\ref{mass2}),
and straightforwardly integrate over $z$ to obtain:
\begin{equation}
\rho(z)=\left[\rho_c-\frac{\alpha\rho_0 H_2}{V_0^{1/3}}
\left(e^{-z/H_2}-1\right)\right]\,e^{-z/H_1}\,,
\label{rhoz}
\end{equation}
where $H_2^{-1}=H^{-1}-2/(3H_1)$. Equation (\ref{rhoz})
agrees with the exact (numerical) integration of
equations (\ref{mass2}, \ref{en2}) to within 2.5\%\ in
the $z=0\to 4H$ range.

We have not, however, found an exact or approximate
integration of Equation (\ref{mom2}), so as to obtain
the vertical velocity $u$ of the thermal as a function
of $z$. Following the suggestion of Nusser et al. (2006),
one can obtain an estimate of the flow velocity by
assuming an approximate balance between the buoyancy
and the drag terms (i.e., setting the right hand
side of equation \ref{mom2} to zero). From this condition,
one finds $u$ as a function of $\rho$, $\rho_a$ and $V$.
Through the previously determined dependencies of
these variables on $z$ one then finds $u(z)$.

However, we find that the vertical velocity $u(z)$ determined
in this way agrees with the exact (numerical)
integration of equation (\ref{mom2}) only to an order of
magnitude. Therefore, a reasonably accurate description
of the rising thermal solution can only be obtained by using
the analytic expressions for the volume $V(z)$ and
the density $\rho(z)$ (equations \ref{volz} and \ref{rhoz})
and then carrying out a numerical integration of
equation (\ref{mom2}) in order to obtain $u(z)$. The
``thermal'' solution as a function of $t$ is then
obtained by integrating the differential equation
$dz/dt=u$.


\begin{thebibliography}

\bibitem{} Blandford, R. D., Rees, M. J. 1974, MNRAS, 169, 395

\bibitem{} Mathews, W. G., Brighenti, F.,
Buote, D. A., Lewis, A. D. 2003, ApJ, 596, 159

\bibitem{} Nusser, A., Silk, J.,
Babul, A. 2006, MNRAS, 373, 739

\bibitem{} Pope, E. C. D., Babul, A.,
Pavlovski, G., Bower, R. G., Dotter, A.
2010, MNRAS, 406, 2023

\bibitem{} Raga, A. C., Cant\'o, J.,
Rodr\'\i guez, L. F., Vel\'azquez, P. F. 2012, MNRAS, 424, 2522

\bibitem{} Raga, A. C., Reipurth, B. 2004, RMxAA, 40, 15

\bibitem{} Raga, A. C., Navarro-Gonz\'alez, R.,
Villagr\'an-Muniz, M., 2000, RMxAA, 36, 67

\bibitem{} Rodr\'\i guez-Gonz\'alez, A., Raga, A. C.,
Cant\'o, J. 2009, A\&A, 501, 411

\bibitem{} Turner, S. 1980, ``Buoyancy effects in fluids'',
Cambridge monographs on mechanics (Cambridge Univ. Press)


\end{thebibliography}
\end{document}